# Low-Field Ferroelectricity in 10 nm AlBScN Thin Films

Xiaolei Tong, Pedram Yousefian, Ziyi Wang, Meenakshi A. Saravanan, Rajeev Kumar Rai, Giovanni Esteves, Eric A. Stach, and Roy H. Olsson III

*Abstract*— Ferroelectric aluminum scandium nitride ($Al_{1-x}Sc_xN$, AlScN) offers CMOS-compatible integration but suffers from high coercive fields and leakage currents that hinder thickness scaling. Further reduction in thickness is essential for low-voltage embedded nonvolatile memory applications. Boron incorporation into AlScN (AlBScN) suppresses leakage current in films down to 40 nm, yet its ferroelectric characteristics in ultrathin films remains unexplored. This letter demonstrates robust ferroelectric switching in 10 nm sputtered AlBScN capacitors with a low coercive field and approximately two orders of magnitude lower leakage than AlScN. Notably, ferroelectric switching was observed at 2.2 MV/cm in capacitance-voltage measurements, and symmetric polarization reversal occurred near 4.6 MV/cm in positive-up-negative-down (PUND) measurements using 2 µs pulses. Moreover, Weibull analysis revealed a breakdown-to-coercive-field ratio ($E_{BD}/E_c$) of ~2.2. These findings demonstrated AlBScN as a promising candidate for CMOS back-end-of-line (BEOL) compatible ferroelectric applications with improved energy consumption and reduced leakage current.

## I. Introduction

FERROELECTRIC materials are widely utilized in non-volatile memory devices, yet many traditional ferroelectrics lack CMOS back-end-of-line compatibility and exhibit degraded ferroelectricity at reduced thicknesses [1], [2]. Aluminum scandium nitride (AlScN) has emerged as a compelling candidate, offering advantages in remnant polarization (>100 µC/cm$^2$), processing temperature (<400 °C), thermal stability (~1000 °C), and thickness scalability (to ~4 nm) [3], [4], [5], [6], [7]. While reducing the thickness of AlScN is essential for lowering the operational voltage in embedded nonvolatile memory applications [5], its coercive field ($E_c$) tends to increase with decreasing thickness due to a higher fraction of the in-plane compressive interface layer [5], [7], [8]. Elevated $E_c$ in ultrathin films exacerbates leakage currents [6], [9], obscuring ferroelectric switching [10], causing read disturb issues [11], and facilitating Joule heating induced failure [12]. Thus, suppressing leakage current is critical for further device scaling, stability, and reliability [11].

Atomic-scale studies indicate that boron incorporation into AlN introduces local structural disorder, lowering the switching energy barrier through a transient non-polar state and thereby reducing leakage current [13]. Building on this, boron and scandium co-doped AlN ($Al_{1-x-y}B_xSc_yN$, AlBScN) has demonstrated substantially lower leakage current at room temperature [9], [14] and up to 600 °C [15].

Our previous study demonstrated ferroelectricity in AlBScN down to 40 nm [15]. Here, we extend this to 10 nm AlBScN, revealing distinct polarization switching with markedly reduced leakage current and lower $E_c$ compared to AlScN of similar thicknesses, establishing AlBScN as a promising energy-efficient and CMOS-compatible ferroelectric material.

## II. Device Fabrication And Characterization

The entire film stack was sputtered in an Evatec CLUSTERLINE 200 II system. A 54 nm Al (111) bottom electrode was deposited at 150 °C on 6-inch c-plane sapphire substrates to reduce in-plane stress while promoting c-axis-oriented growth [5]. AlBScN layer were co-sputtered from 100 mm $Al_{0.89}B_{0.11}$ and Sc targets operated at 900 W and 655 W, respectively, with $N_2$/Ar flows of 30/8 sccm at 300 °C, and in-situ capped with Al to prevent oxidation [5]. Top electrodes (5, 10, 20 µm diameters) were defined by laser lithography. To enable probing of small capacitors, a 70 nm $SiO_2$ layer with vias was deposited, followed by patterning of 100 nm Al contact pads with 50 µm diameter using the same lithography process.

Structural characterization was performed using a Rigaku SmartLab SE X-ray diffractometer (XRD), a JEOL F200 transmission electron microscope (TEM) and an aberration-corrected JEOL NEOARM TEM operating at 200 kV. Electrical measurements were conducted using a Keithley 4200A-SCS on 10 µm diameter capacitors. Ferroelectric switching was characterized by capacitance–voltage sweep and

Manuscript received X; revised X; accepted X. This work is supported in part by the Defense Advanced Research Project Agency under the COFFEE program and in part by Army/ARL via the Collaborative for Hierarchical Agile and Responsive Materials (CHARM) under cooperative agreement W911NF-19-2-0119. The deposition, patterning, and characterizations of $Al_{1-x-y}B_xSc_yN$ thin films were performed at the Singh Center for Nanotechnology, which is funded under the NSF National Nanotechnology Coordinated Infrastructure Program (NNCI-1542153). The authors acknowledge the use of an XRD facility supported by the Laboratory for Research on the Structure of Matter and the NSF through the University of Pennsylvania Materials Research Science and Engineering Center (MRSEC) DMR2309043.

X. Tong, P. Yousefian, Z. Wang, M. A. Saravanan, and R. H. Olsson III are with the Department of Electrical and Systems Engineering, University of Pennsylvania, USA (e-mail: rolsson@seas.upenn.edu).
R. K. Rai, and E. A. Stach are with the Department of Materials Science and Engineering, University of Pennsylvania, USA.
G. Esteves is with Microsystems Engineering, Science and Applications, Sandia National Laboratories, Albuquerque, New Mexico, USA.

positive–up–negative–down (PUND) measurements (2 µs pulses, 300 ns rise/fall, 500 ns delay). Leakage current was measured by DC I-V sweeps. $E_{BD}$ was determined using 10 kHz triangular wave hysteresis measurements on a Radiant Precision Premier II, and Weibull statistics were applied to extract the breakdown field at 63.2% failure probability.

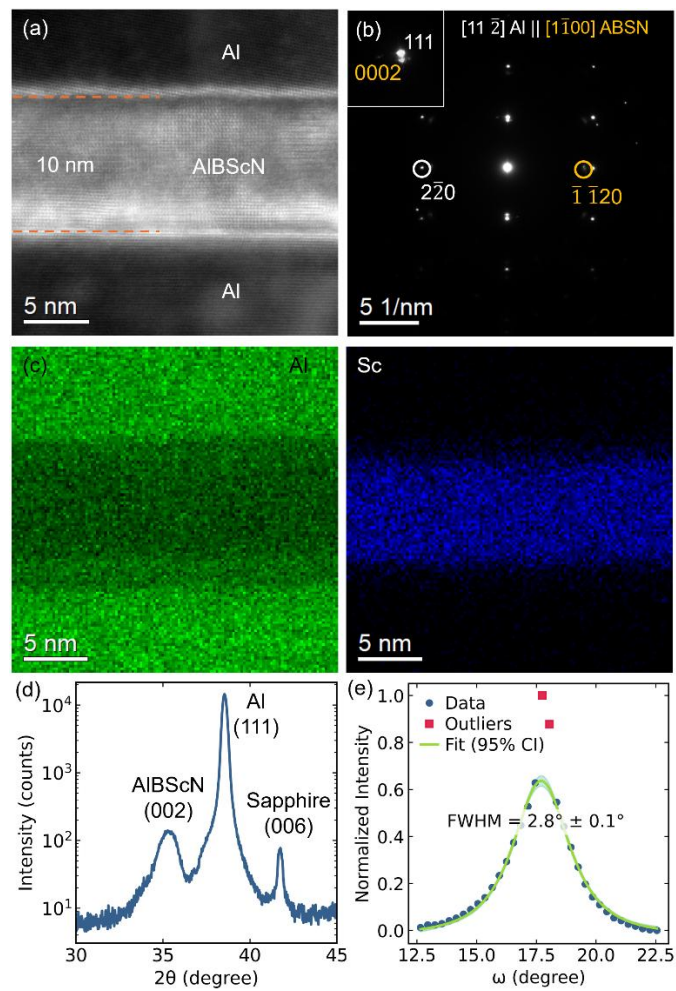

Fig. 1. (a) Cross-sectional TEM image of the Al/AlBScN/Al, (b) Selected area electron diffraction (SAED) pattern, (c) Energy dispersive X-ray spectroscopy (EDS) of the film stack, (d) XRD θ–2θ measurement, and (e) rocking curve measurement fitted by Pearson VII profile with 95% confidence interval. The outlier points correspond to an unrealistically narrow feature (FWHM ≈ 0.5°) and are therefore not attributed to the AlBScN (002) reflection. These points do not significantly affect the fitted peak shape or extracted FWHM.

## III. EXPERIMENTAL RESULTS

TEM cross-sectional imaging and energy dispersive X-ray spectroscopy (EDS) mapping confirmed a 10 nm AlBScN film sandwiched between highly crystalline Al layers with uniform distribution of Al and Sc ((Fig. 1 (a) and (c)). Detection of boron by EDS was hindered by its low-energy $K_\alpha$ emission (~0.183 keV), which suffers from strong absorption, poor detector efficiency, and low fluorescence yield. Selected area electron diffraction (SAED) pattern revealed a (002) textured wurtzite AlBScN film epitaxially grown on Al (111) bottom electrode (Fig. 1(b)), consistent with the distinct AlBScN (002) reflection along with Al (111) peaks observed in XRD θ–2θ scans (Fig. 1(d)). Rocking curve measurements of the AlBScN (002) reflection show a peak with a full width at half maximum (FWHM) of 2.8° (Fig. 1 (e)), suggesting improved crystalline quality compared with our previous ~10 nm AlScN on Al and 40 nm AlBScN on Pt bottom electrodes [5], [15].

Ferroelectric switching was confirmed by the butterfly-shaped C–V loop (Fig. 2 (a)) with $E_c$ of 2.2 MV/cm, lower than the reported ~2.8 and ~3.1 MV/cm for ~10 nm AlScN measured by the same method [16], [17]. PUND measurements further illustrated the ferroelectric switching evolution in Fig. 2(d)–(g). The 3.0 V pulse amplitude trace showed no measurable switching current, whereas distinct current differences appeared in the N–D and P–U pulses at 4.0 V and 4.6 V, indicating ferroelectric switching with an earlier onset in the N pulse, consistent with imprint behavior in AlScN at similar thickness [17]. At 5.6 V, both P and N pulses exhibited clear switching currents (Fig. 2(h)). Extracting accurate $2P_r$ values was not feasible because the switching current could not be sufficiently isolated from the capacitive and leakage components at this ultrathin thickness. Nevertheless, the extracted $E_c$ of 4.6 MV/cm under pulsed conditions was lower than previously reported values for AlBScN with thickness ranging from 40 to 269 nm measured by PUND and polarization hysteresis [14], [15], and comparable to that reported in 250 nm AlBScN films [9]. It was also lower than values reported for AlScN films of comparable thickness [5], [8], [16], [17], [18], [19]. These results demonstrate the potential for further thickness scaling of AlBScN on Al without increasing $E_c$, thereby enabling lower switching voltages required for CMOS applications.

The $E_{BD}/E_c$ ratio defines the operational voltage window for stable and reliable polarization switching without dielectric failure. Breakdown measurements showed increasing $E_{BD}$ and narrower Weibull distributions with decreasing device area (Fig. 2(b)), consistent with defect-limited breakdown behavior observed in AlScN [5], [20]. The characteristic $E_{BD}$ at 63.2% failure probability was 8.8, 10.0, and 10.1 MV/cm for 20, 10, and 5 µm capacitors, respectively, yielding $E_{BD}/E_c$ ratio of ~2.2 for the 10 µm diameter capacitors.

DC I–V sweeps were conducted to determine whether the leakage suppression effect of boron incorporation is retained upon thickness scaling. The 10 nm AlBScN capacitors exhibited approximately two orders of magnitude lower leakage than 10 nm AlScN [5] at the same electric field, indicating that boron effectively suppressed leakage even in ultrathin films (Fig. 2(c)). This finding confirmed that the leakage reduction observed in thicker AlBScN layers [9], [14], [15] persisted at 10 nm, underscoring the scalability of AlBScN for CMOS-compatible ferroelectric integration.

## IV. CONCLUSION

In this study, 10 nm AlBScN capacitors were fabricated with confirmed c-axis-oriented wurtzite film structure. Electrical characterizations showed reduced leakage current [5] and lower $E_c$ compared to AlScN at similar thicknesses (Table I), highlighting AlBScN as a promising CMOS compatible ferroelectric material.

### ACKNOWLEDGEMENT

This article has been authored by an employee of National Technology & Engineering Solutions of Sandia, LLC under Contract No. DE-NA0003525 with the U.S. Department of Energy (DOE). The employee owns all right, title and interest in and to the article and is solely responsible for its contents. The United States Government retains and the publisher, by accepting the article for publication, acknowledges that the United States Government retains a non-exclusive, paid-up, irrevocable, world-wide license to publish or reproduce the published form of this article or allow others to do so, for United States Government purposes. The DOE will provide public access to these results of federally sponsored research in accordance with the DOE Public Access Plan. https://www.energy.gov/downloads/doe-public-access-plan.

TABLE I

COMPARISON OF $E_c$ IN ~10 NM ALSCN FILMS. ALL WERE DEPOSITED BY SPUTTERING EXCEPT [16], WHICH UTILIZED MOLECULAR BEAM EPITAXY.

| Bottom electrode/Material | $E_c$ (MV/cm) | Measurement Method |
|---|---|---|
| Pt / 9 nm $Al_{0.8}Sc_{0.2}N$ [8] | 9.8 | PUND (10 µs) |
| Pt / 10 nm $Al_{0.72}Sc_{0.28}N$ [17] | ~3.1 | C–V |
|  | ~6.3 | J–E loop (80 kHz) |
| Al / 9.6 nm $Al_{0.72}Sc_{0.28}N$ [5] | 9.0 | PUND (500 ns) |
| Mo / 12 nm $Al_{0.7}Sc_{0.3}N$ [16] | ~2.8 | C–V |
|  | ~5.1 | PUND (20 kHz) |
| $HfN_{0.4}$ / 10 nm $Al_{0.7}Sc_{0.3}N$ [7] | ~8.5 | PUND (100 kHz) |
| Al / 10 nm AlBScN [this work] | 2.2 | C–V |
|  | 4.6 | PUND (2 µs) |

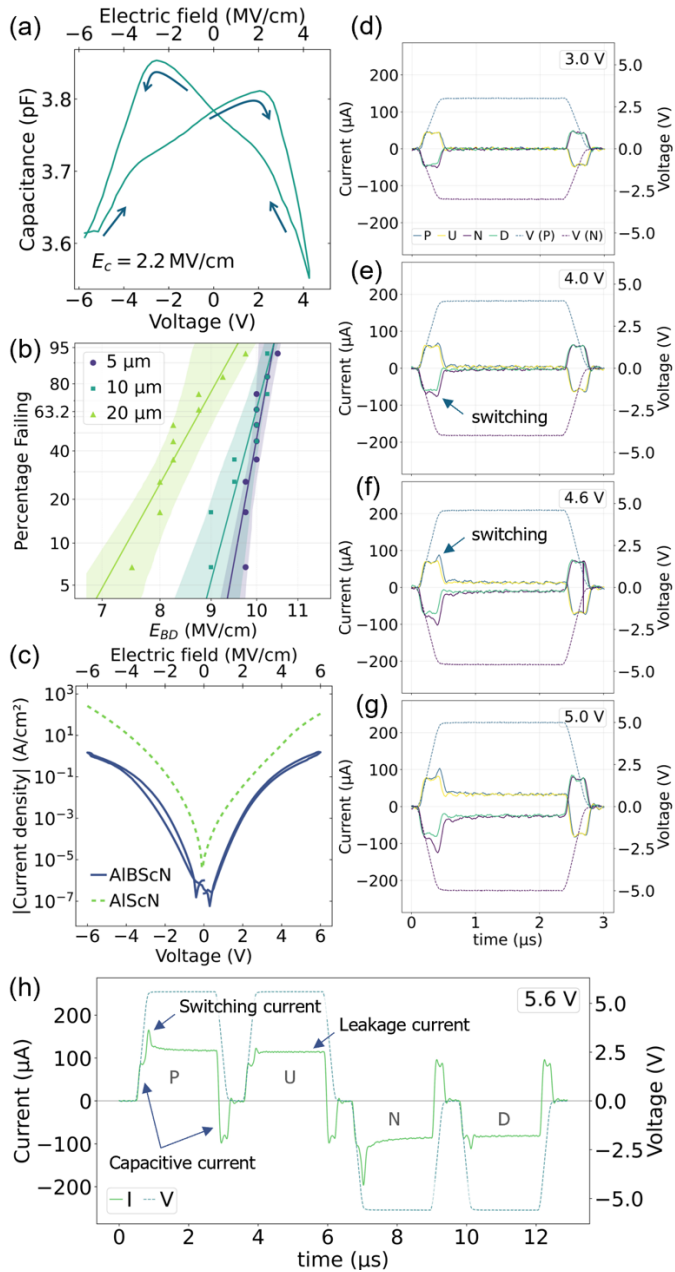

Fig. 2 (a) C-V loop, (b) Weibull distribution plot of $E_{BD}$, (c) DC-IV measurement, (d)-(g) Overlayed current response from PUND, (h) Fulltime PUND measurement at 5.6V applied voltage.